# Agent-Based Simulation Modelling for Reflecting on Consequences of Digital Mental Health

A working paper based on the 2018 Summer Internship Report by Daniel Stroud, who was supervised by Peer-Olaf Siebers and Christian Wagner

Finalised by Peer-Olaf Siebers (University of Nottingham, School of Computer Science, Nottingham, NG8 1BB, UK, peer-olaf.siebers@nottingham.ac.uk)

## Abstract
The premise of this working paper is based around agent-based simulation models and how to go about creating them from given incomplete information. Agent-based simulations are stochastic simulations that revolve around groups of agents that each have their own characteristics and can make decisions. Such simulations can be used to emulate real life situations and to create hypothetical situations without the need for real-world testing prior. Here we describe the development of an agent-based simulation model for studying future digital mental health scenarios. An incomplete conceptual model has been used as the basis for this development. To define differences in responses to stimuli we employed fuzzy decision making logic. The model has been implemented but not been used for structured experimentation yet. This is planned as our next step.

## 1 Introduction

Agent-Based Simulation (ABS) is a powerful paradigm that can be used for conducting what-if analysis of human centric systems (Siebers and Aickelin 2008). Agents are comparable to non-player characters in games. In principle, ABS enables exploring the interaction of different groups of stakeholders, where the latter are often people in a specific context, from land owners in conservation, to shoppers in marketing.

A key challenge in ABS is the appropriate specification of agents in order for them to behave similarly to humans. Commonly, so-called archetypes are established, which capture groups within a population sharing similar behaviour (Zhang et al 2012). This project is designed to combine fuzzy set theory with ABS in order to enable simulations where agents can be efficiently designed to replicate core behavioural aspects of human stakeholders, thus providing a pathway for human-inspired ABS.

My internship was based on creating a simulation from an incomplete conceptual model that was previously created by my supervisor Peer-Olaf Siebers and his colleagues Elvira Perez Vallejos and Tommy Nilsson (Siebers et al unpublished) using the Engineering Agent Based Social Simulation framework (or EABSS for short). The EABSS supports a structured way of co-creation by embedding design philosophies and tools from Software Engineering and using these for driving its focus group communications. The process is guided by a set of predefined table templates in combination with the use of UML diagrams - a graphical notation used in Software Engineering to conduct system analysis and design (Fowler 2004) - for capturing the thoughts of academic and non-academic participants.

I started by learning to use AnyLogic (XJTEK 2018) and by reading up on the subject of technology with mental health. After this, I created a very simple simulation of a hospital scenario including doctors, patient and robot doctors (the technology). This simulation gave me something that I could add to and build upon using the conceptual model given to me by Peer. I continued adding to the conceptual model and the simulation model until I had a detailed enough simulation that I could start catering it towards the hypotheses defined in the conceptual model I was given. As well as this, I added a small bit of fuzzy logic to the simulation to familiarize myself with the area before the second half of my internship. The fuzzy logic was used as the decision making for doctors and robot doctors to decide how long they should treat patients for, and was used for visitors to decide whether they should visit or not. After catering towards the hypotheses I had a fully working simulation based off the conceptual model given to me.

## 2 Proof-of-Principle Agent-Based Simulation Model

For the first half I created an ABS based on a conceptual model given to me by Peer Olaf Siebers that was created in a focus group (Siebers et al, unpublished). In Agent-Based Modelling (ABM), a system is modelled as a collection of autonomous decision-making entities called agents. Each agent individually assesses its situation and makes decisions on the basis of a set of rules (Bonabeau 2002). He wanted me to create a simple model that would visualize and give valuable outputs to the problems posed in the conceptual model. Some scans of the original conceptual model are shown in Figure 1. The conceptual model had the aim "reflect on consequences of digital mental health solutions" and used the EABSS framework approach (Siebers and Klügl 2017) for driving the model development process. The information within it allowed me to develop a simulation model and run some experiments. My simulation was a hospital scenario with doctors, patients, visitors and robot doctors (modelled as proactive machines). The idea of the hospital scenario was to have doctors and proactive machines treating patients with occasional visits from visitors and then to track the mental state of the patients. Depending on the different ratios of human doctors to proactive machines and how much the patients communicated, their mental states would vary.

Aim
- Reflect on consequences of digital mental health solutions {consider relationship between machines and humans}

Objectives
- Investigate how much control we are willing to give up for convenience {"trust"}
- What is the impact {in what respect?} of receiving service at home 24/7

Hypotheses
- Privacy is predominant (trade away privacy for convenience)
- Group pressure makes people argue that there is a need for more privacy
- Trust between machines and humans will affect human-human relationships in a negative way {"apart"}
- Machine-look influences our decisions (trust)
- Lower costs of robots leads to increase in demand
- Group pressure leads to robot uptake (trust)

*Handwritten annotations:* human/human relationship, human/machine; To investigate; Does culture mediate relationship with technology; economic aspect, don't want to be behind; Technology promotes loneliness; mobile phone bus example

Experimental factors
- Initial number of people receiving service by machines at home
- Initial costs of robots

*Handwritten:* Privacy policy – Attitude towards robot

Responses
- Final number of people receiving service by machines at home
- Final costs of robots
- Number of marriages between machines and humans

*Handwritten:* Connectivity < online/offline } Network structure

Purpose
- Transparency

EABSS driven system analysis

| | | Include | Exclude | Group | Justification |
|---|---|---|---|---|---|
| **Actors** | 1 Intelligent machines | | | 1+2: Proactive machines | |
| | 2 Human like machines | | | 1+2: Proactive machines | |
| | 3 Patients | | | | |
| | 4 Developers of machines | | | | |
| | 5 Clinicians | | | | |
| | 6 Service providers (collecting personal data) | | | | |
| | 7 Pets | | | | |
| | 8 Policy makers | | | 8+9: Rule makers | |
| | 9 Regulators | | | 8+9: Rule makers | |
| | 10 Families | | | 10+11: Families | |
| | 11 Couples | | | 10+11: Families | |
| | 12 Singles | | | | |
| **Physical Env.** | 13 Home (private and public areas) | | | | |
| | 14 Clinic | | | | |
| | 15 Workspace | | | | |
| | 16 Public space | | | | |
| | 17 Mobile phone (handheld device) | | | | |
| | 18 Tablet | | | | |
| | 19 Laptop | | | | |
| | 20 Weather | | | | |
| **Social and Psychol. Aspects** | 21 Influence of anger and fear on ethical decisions | | | | |
| | 22 Trust | | | | |
| | 23 Loneliness | | | | |
| | 24 Human-Computer Interaction | | | | |
| | 25 Severity of mental health symptoms | | | | |
| | 26 Theory of planned behaviour | | | | |
| | 27 Resilience | | | | |
| | 28 Privacy / Anonymity | | | | |
| | 29 Child trauma (how to interact with intelligent machines) | | | | |
| **Other** | 30 Networks (social, media, peer support) | | | | |
| | 31 | | | | |
| | 32 | | | | |
| | 33 | | | | |

EABSS driven initial model scope table

| | | | Include | Exclude | Group | Justification |
|---|---|---|---|---|---|---|
| Actors | 1 | Intelligent machines | ✓ | | 1+2: Proactive machines | |
| | 2 | Human like machines | ✓ | | 1+2: Proactive machines | |
| | 3 | Patients | | | | |
| | 4 | (Developers of machines) | | | | |
| | 5 | Clinicians | | | | |
| | 6 | Service providers (collecting personal data) | | | | |
| | 7 | ~~Pets~~ | | | | |
| | 8 | Policy makers | ✓ | | 8+9: Rule makers | |
| | 9 | Regulators | ✓ | | 8+9: Rule makers | |
| | 10 | Families | ✓ | | 10+11: Families | |
| | 11 | Couples | ✓ | | 10+11: Families | |
| | 12 | ~~Singles~~ Non-patient | | | | |
| Physical Env. | 13 | Home (private and public areas) | | | | |
| | 14 | Clinic | | | | |
| | 15 | ~~Workspace~~ Private space | | | | |
| | 16 | Public space | | | | |
| | 17 | Mobile phone (handheld device) } exist or not | | | | |
| | 18 | Tablet | | | | |
| | 19 | Laptop | | | | |
| | 20 | ~~Weather~~ | | | | |
| Social and Psychol. Aspects | 21 | Influence of anger and fear on ethical decisions | | | | |
| | 22 | Trust | | | | |
| | 23 | Loneliness | | | | |
| | 24 | Human-Computer Interaction | | | | |
| | 25 | Severity of mental health symptoms (how to treat people) | | | | |
| | 26 | Theory of planned behaviour | | | | |
| | 27 | Resilience | | | | |
| | 28 | Privacy / Anonymity / Security | | | | |
| | 29 | Child trauma (how to interact with intelligent machines) | | | | |
| Other | 30 | Networks (social; ~~media~~; peer support) fashion → trends cultural | | | | |
| | 31 | | | | | |
| | 32 | | | | | |
| | 33 | | | | | |

EABSS model scope table after iteration

Figure 1: Scans of parts of the original conceptual model that I worked off

The initial steps of the internship were to create a very simple hospital scenario that wasn't very specific to certain hypotheses, this would allow me to add more detail later depending on which hypotheses from the conceptual model I chose to focus on. My initial model included doctors, proactive machines and patients. The doctors and proactive machines would both treat the patients and there was no mental state for the patients at this point. They treated the patients based off a linked list containing patients that needed a check-up or requested to be seen. For the implementation I used the free version of AnyLogic PLE (XJTEK 2018). Figure 2 shows how the simulation looked like during runtime at this point (stage 1).

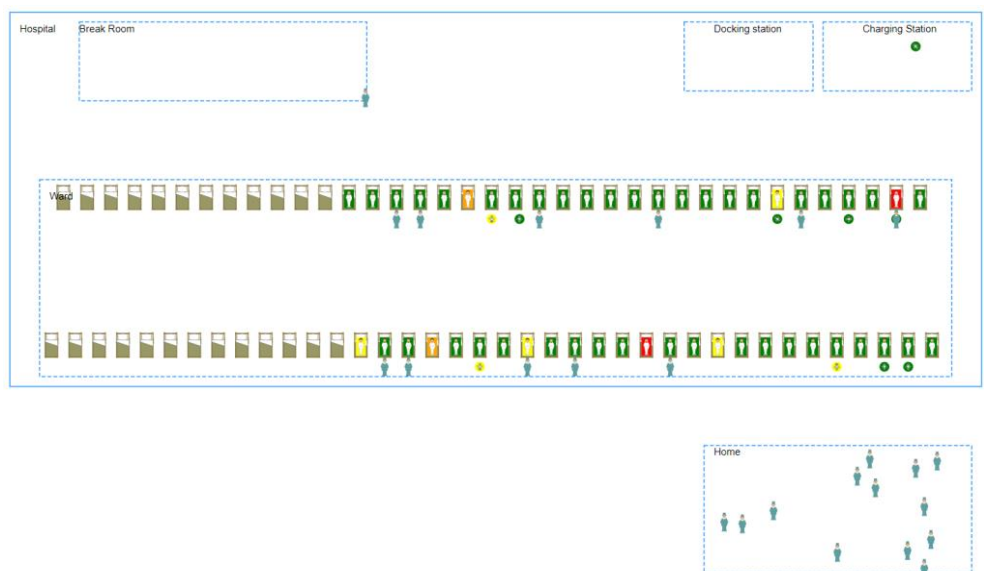

Figure 2: Example screenshot, showing what the simulation looked like at stage 1

The next step after this was adding visitors and differences between the doctors and proactive machines. I initially added visitors that would be created at a certain time every day and then after an hour they would all be removed from the simulation. This method caused some problems due to the 50,000 agent limit on the free version of AnyLogic, so I had to create a certain number of visitors at the start of the simulation and then have them visit and leave the hospital every day for the duration of the simulation. In order to create differences between the doctors and proactive machines, firstly I made them do slightly different jobs and secondly, I added a mental state (satisfaction) variable to the patients so that they are more satisfied if they have been treated by a doctor than a proactive machine. The colour behind each patient represents their mental state. Figure 3 shows how the simulation looked like during runtime at this point (stage 2).

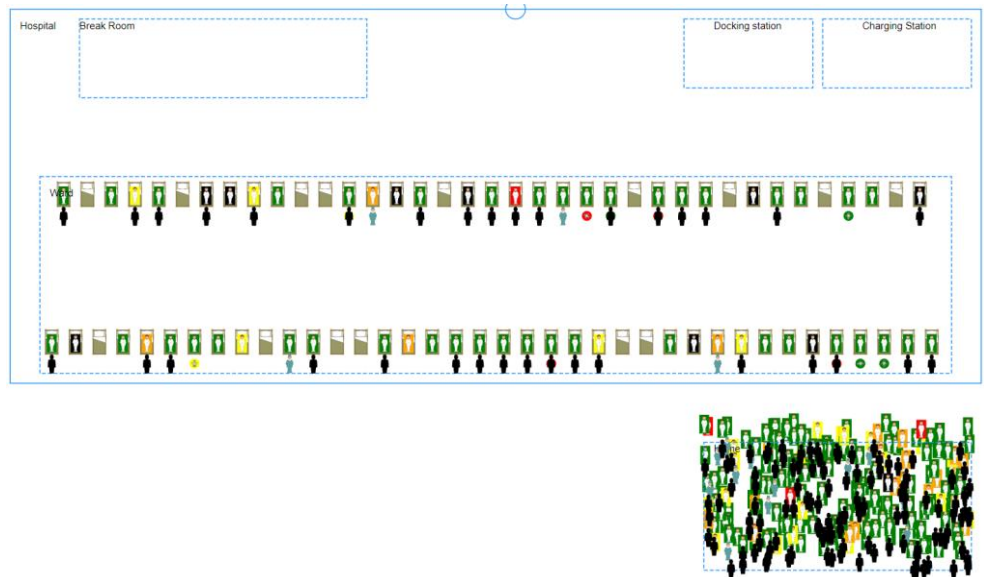

Figure 3: Example screenshot, showing what the simulation looked like at stage 2

## 3 Proof-of-Principle Fuzzy Logic Extension

At this point I had a simulation that was fully functional and allowed me to track the patients' mental states over time, however I had not focussed on certain hypotheses from the original conceptual model.

Before fully focussing on specific hypotheses I decided to create and add some fuzzy logic systems into the simulation to give me a better idea of the area for the second half of the internship and because some areas of the simulation required decision making that could be made using fuzzy logic systems. Fuzzy logic is an approach to computing based on "degrees of truth" rather than the usual "true or false" (1 or 0) Boolean logic on which the modern computer is based. The idea of fuzzy logic was first advanced by Dr Lotfi Zadeh of the University of California at Berkeley in the 1960s (Rouse n.d.). I added three FLSs overall and their purposes were to decide the amount of time doctors and proactive machines should treat patients for and also to decide whether visitors should visit patients and if so, how long for. I created these FLSs initially using JuzzyOnline (Wagner and Pierfitte n.d.) and then used those to help write and implement the Java into the AnyLogic simulation model I created earlier. Following are the fuzzy logic graphs showing the inputs, outputs and membership functions of the FLS' for a doctor (Figure 4), a robot (Figure 5), and a visitor (Figure 6).

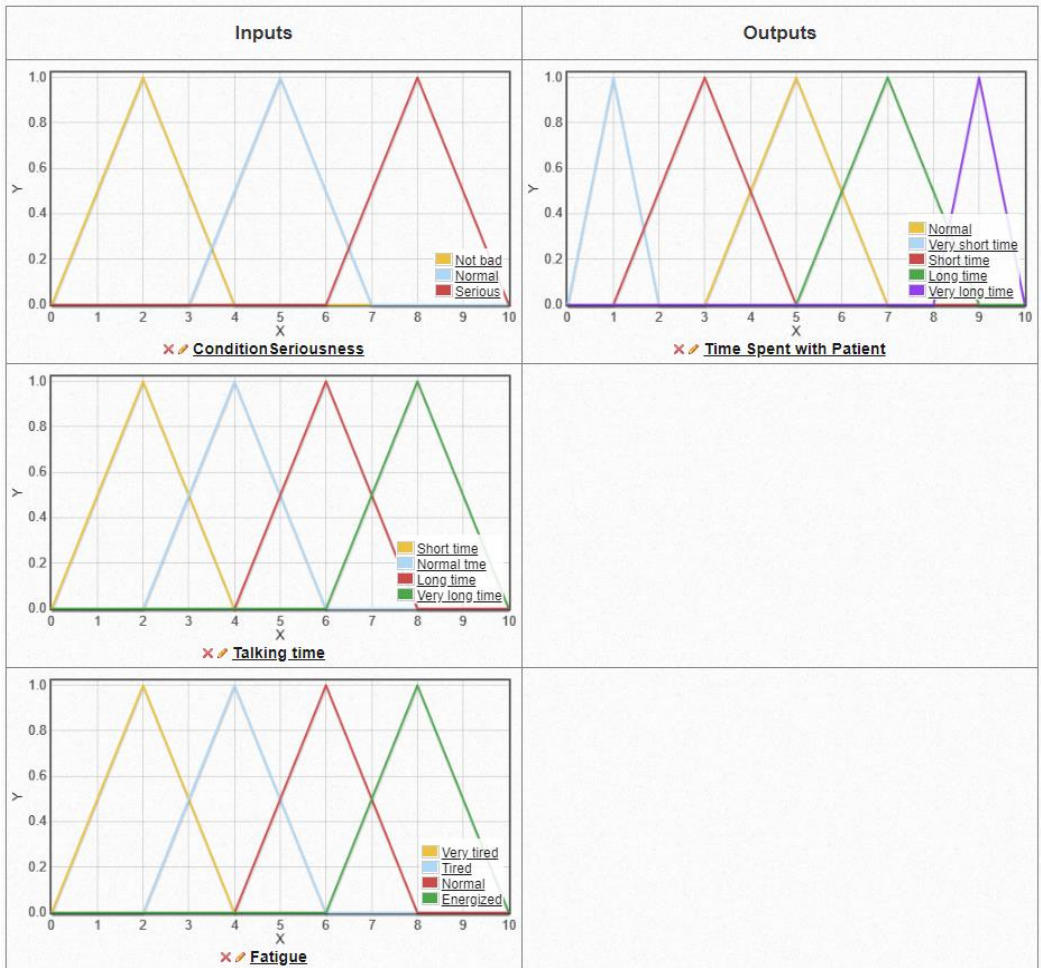

Figure 4: Inputs, outputs and membership functions of the FLS' for a doctor

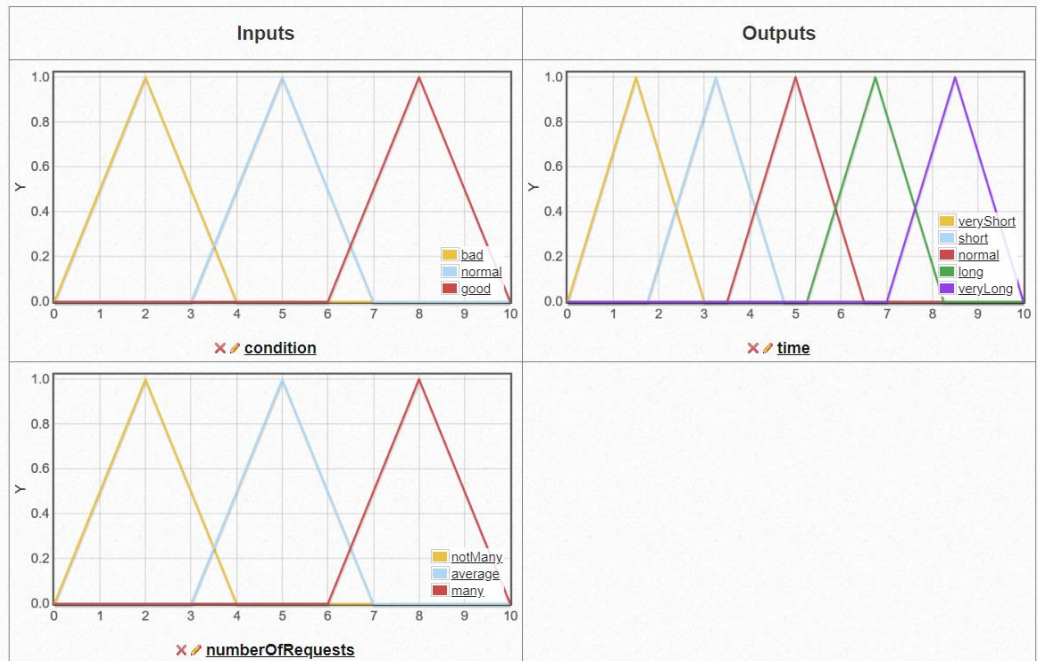

Figure 5: Inputs, outputs and membership functions of the FLS' for a robot

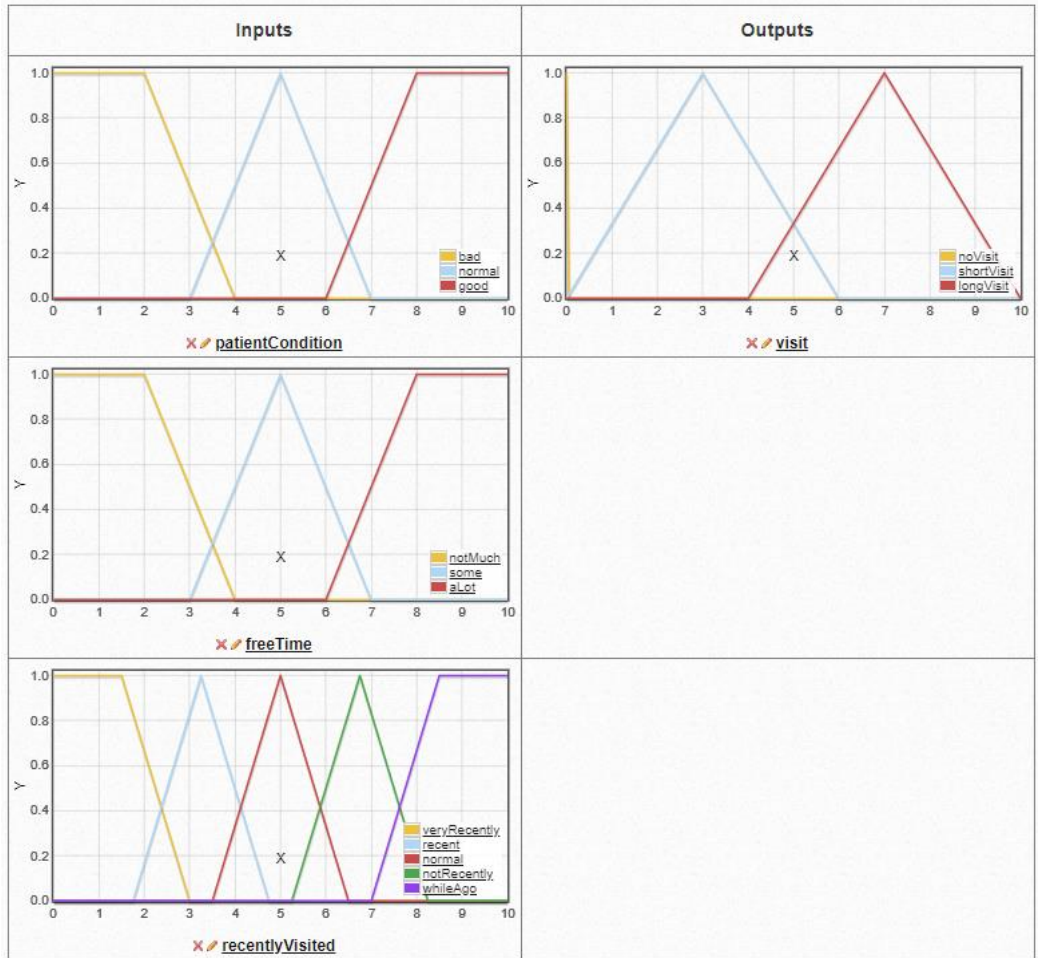
Figure 6: Inputs, outputs and membership functions of the FLS' for a visitor

# 4 Applying EABSS for Model Re-Development and Documentation

After adding fuzzy logic into the simulation I decided to focus on 2 of the hypotheses posed by the original conceptual model. The two hypotheses I focused on were 'Trust between machines and humans will affect human-human relationships in a negative way' and 'Machine-look influences our decisions'.

To aim the simulation towards the first hypothesis I had to add trust variables to each patient. These variables changed over time depending on the treatment of robots or doctors on that patient. As well as this I created a network between the patients so that a patient is connected to the patients in the beds next to them and the beds next to those beds. If patients are connected in a network, then their current trust can affect other patient's trust as if they are talking to each other and sharing opinions. Then to be able to view the relationships/trust between patients they are connected with lines that change colour depending on the patients' difference in trust values. If there is a large difference, then the line is shown in red but if they have similar trust values then they are like each other and trust each other so the line is shown in green. Anywhere in between is shown in yellow.

To tackle the second hypothesis, I created two different groups of robots, human-like robots and robot-like robots. If a human-like robot treats a patient, then they are more satisfied and trust the robot more than if is robot-like. From this I can track the trust and opinions of robots and see it spread throughout the patients using the network explained before. As well as this I added two groups of doctors, a senior group and a junior group. The senior doctors treat patients quicker than the junior doctors meaning that they can treat more patients overall.

In the following I present the conceptual model for final version of hospital simulation using the EABSS framework:

## Title
Exploratory Study of Proactive Machines in a Hospital Scenario

## Context
Proactive machines are an idea that in the future hospitals will have automated machines as well as Doctors and Nurses that can help and treat patients. This project uses the idea of proactive machines to test the impact of technology on mental state and opinions within patients. The project also incorporates the use of fuzzy logic for some of the decision making.

## Gathering Knowledge
All information for this project has been gathered from a conceptual model created by a focus group lead by Peer-Olaf Siebers which had the aim of reflecting on the consequences of digital mental health solutions considering the relationship between machines and humans.

## Step 1: Defining Objectives

### Aim and objectives
Reflect on the consequences of digital mental health solutions

### Hypotheses
- Trust between machines and humans will affect human-human relationships in a negative way
- Machine-look influences our decisions

### Experimental factors
- Number of doctors, proactive machines and patients
- Number of beds

### Responses
- Mental state of patients, e.g. happiness
- Opinions of each patient on the doctors and robots
- The trust of the patients on the robots

## Step 2: Defining the Scope

### Level of abstraction
For this model the high level scope decision was to only consider the treatment of patients by doctors and proactive machines and to consider the visitors visiting patients. Any movement of agents or geographical location were not considered as there is no need for this and it would add more complexity which is not needed. I have chosen to use a hypothetical hospital setting to keep it realistic without having the problems of gathering data which would be required if a real-world hospital was used.

*Scope table*

The next step was to define the scope of the simulation model, considering which elements should be included/excluded and stating why I have made this decision. The result can be found in Table 1.

| Category | | Element | Decision | Justification |
|---|---|---|---|---|
| Actor | Human | Patient | Include | Patients are the ones who are being tested |
| | | Doctor | Include | Doctors need to be compared against the Robot doctors |
| | | Visitor | Include | Visitors can be used to influence patients |
| | | Nurse | Exclude | Nurse can be viewed as the same as doctors in the simulation |
| | Proactive machine | Robot doctor | Include | Need to be compared against doctors |
| Physical Environment | Buildings | Hospital | Include | Location where the model is held |
| | | Home | Include | Location where agents are held when they are being used |
| | Furniture | Beds | Include | Patients need somewhere to stay |
| | Structure | Walls | Exclude | Not needed because we are not interested in agents movements or collisions |
| | | Doors | Exclude | Not needed to test the hypotheses |

Table 1: Scope table (part 1/2)

| Category | | Element | Decision | Justification |
|---|---|---|---|---|
| Social and psychological aspects | Robot doctor behaviour | FLS to decide how long to treat patient for | Include | Needed for more accurate results |
| | | Accurate movement of robot | Exclude | Movement is not needed to test the hypotheses |
| | Doctor behaviour | FLS to decide how long to treat patient for | Include | Needed for more accurate results |
| | | Accurate movement of doctor | Exclude | Movement is not needed to test the hypotheses |
| | Patient behaviour | Random hospital visits | Include | Needed for more accurate results |
| | Visitor behaviour | FLS to decide whether the visitor should visit the patient | Include | Needed for more accurate results |
| | | Accurate movement of visitor | Exclude | Movement is not needed to test the hypotheses |
| Other | | N/A | N/A | N/A |

Table 1: Scope table (part 2/2)

## Step 3: Defining Key Activities

For this stage, use case diagrams were made to show the specific use cases that each actor must do. Figure 7 shows the use case diagram from the original conceptual model and includes a lot of actors that are not in the final simulation. Figure 8 shows the use case diagram from the final simulation that includes just the actors and use cases that are included in the final simulation.

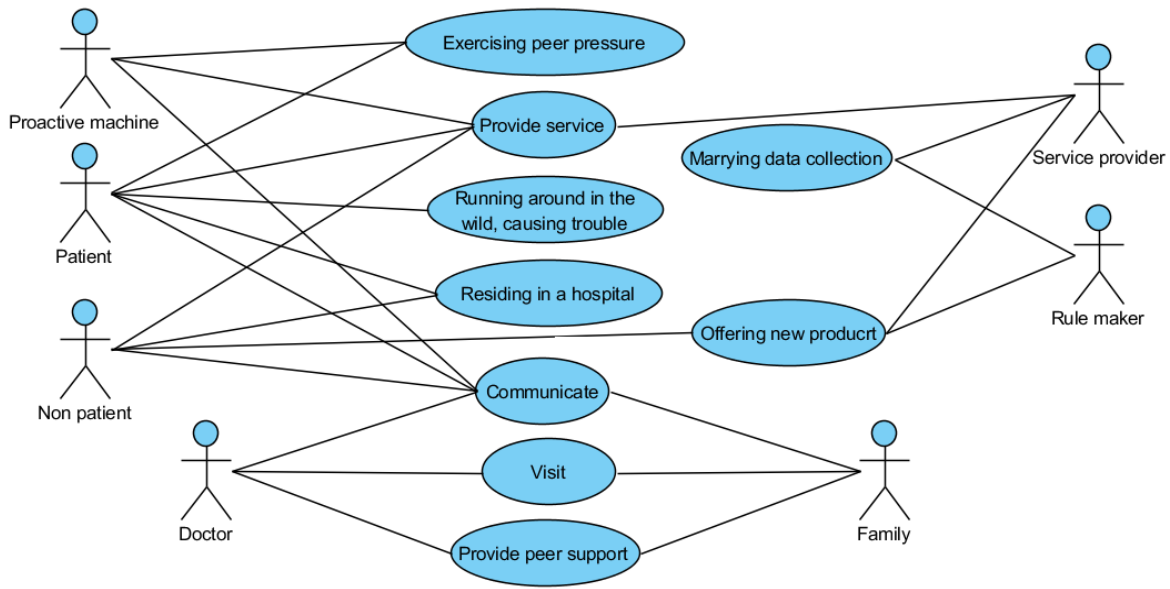

Figure 7: Use case diagram of original conceptual model

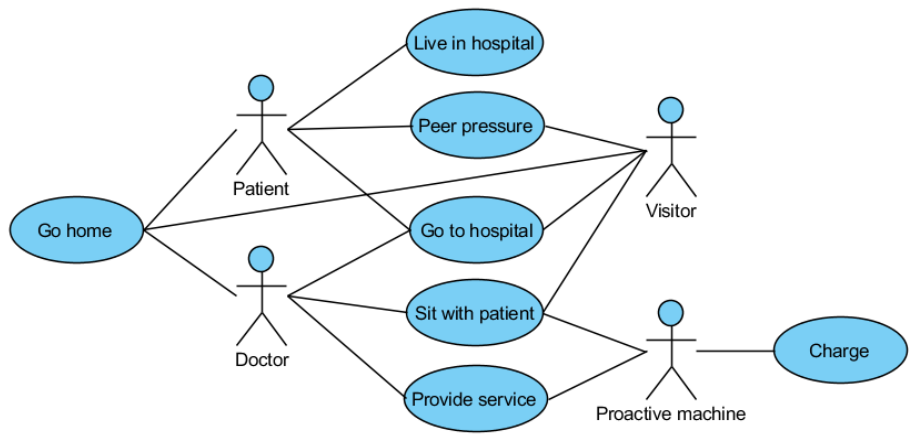

Figure 8: Use case diagram of final simulation

## Step 4: Defining Stereotypes

I decided to use a habit table approach to create the stereotypes. Stereotypes were only needed for the doctors and robots because these were the only agents where stereotyping would have been relevant to the hypotheses. Table 2 shows that a doctor can be one of two stereotypes, a senior or junior. This difference in level means that the senior doctors treat patients quicker than the junior doctors. Table 3 shows that robots can also be one of two stereotypes, they can either be more humanlike in appearance or more robot-like. Whichever one they are, affects the patients opinions on them when they are treated by them.

| Stereotype | Extra time treating patients (Minutes) |
|---|---|
| Senior | 0 |
| Junior | 10 |

Table 2: Doctor stereotypes

| Stereotype | Humanlike variable | Effect on human opinion |
|---|---|---|
| Humanlike | 0.5 – 1 | Positive effect |
| Robotlike | 0 – 0.5 | Negative effect |

Table 3: Robot stereotypes

## Step 5: Defining Agent and Object Templates

*Class diagram*

Figure 9 shows a class diagram for the whole simulation. It shows each agent/class and what is contained inside them as well as how each agent/class is connected to each other. The solid black diamonds represent strong aggregation also known as composition.

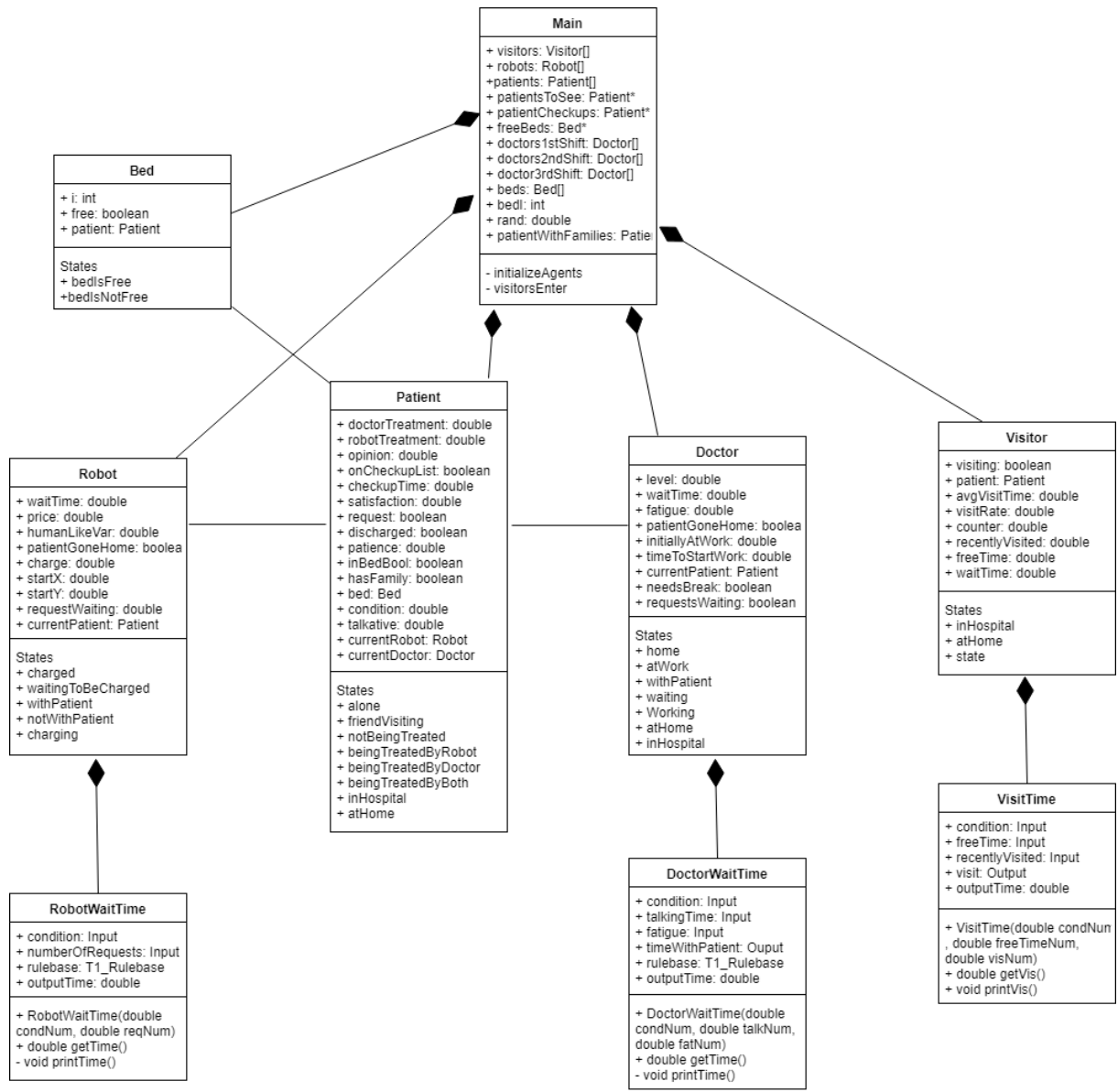

Figure 9: Class diagram for final simulation

*State machine diagrams*

The figures below (Figure 10-14) show the state machine diagrams for each of the agents. Most agents have very simple state machines with usually less than 3 main states. They are mostly self-explanatory with names on each of the states that describe what they do. I have made use of three different types of transitions which are timeout transitions, condition transitions and message transitions. Timeout transitions trigger after a certain amount of time, condition transitions trigger when a condition is true and message transitions trigger when a message is received.

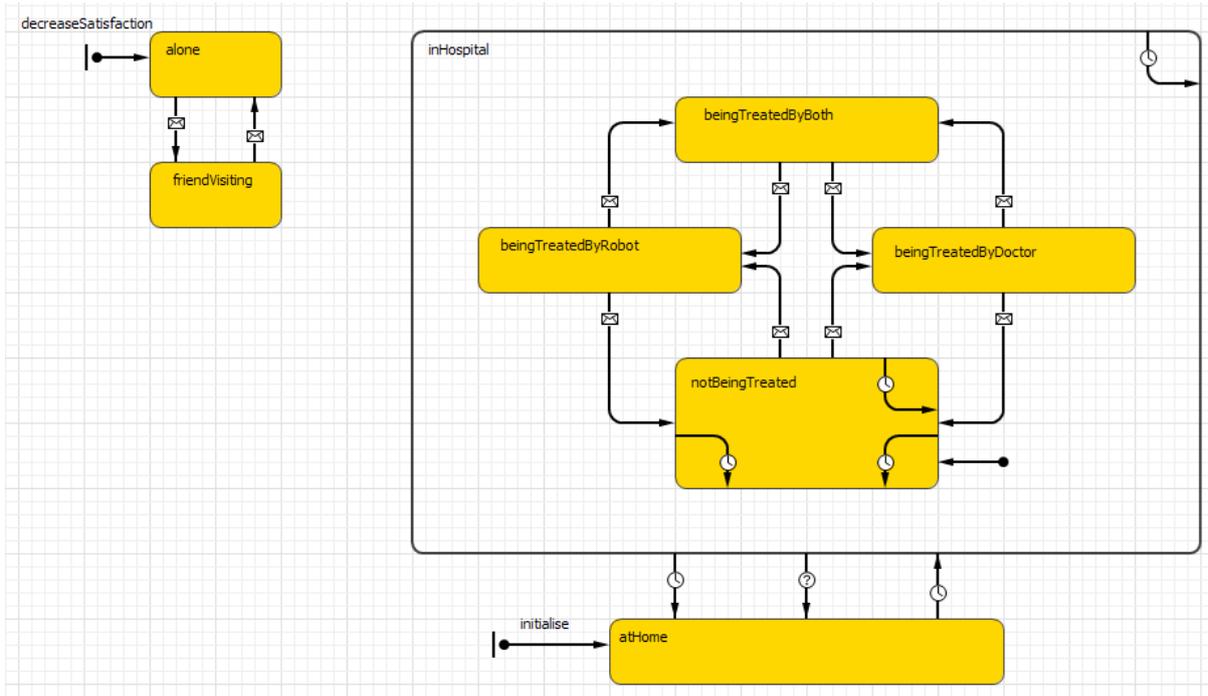

Figure 10: Patient agent

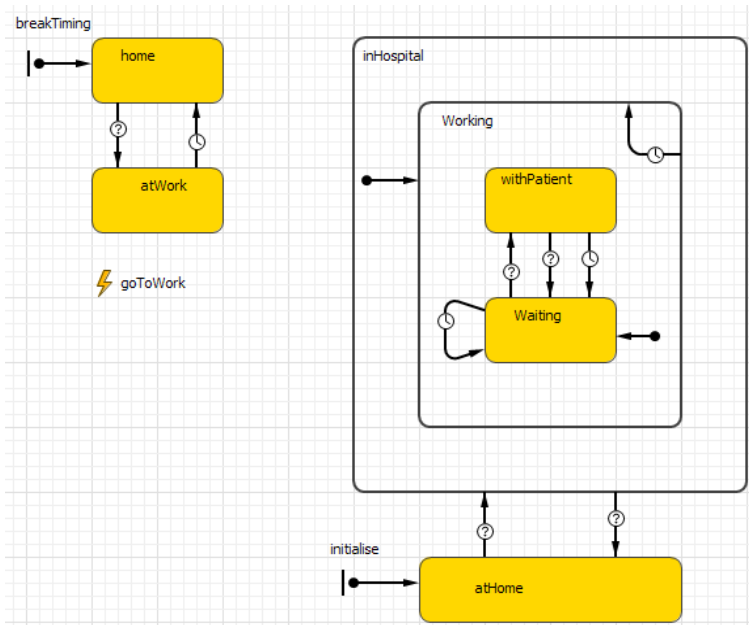

Figure 11: Doctor agent

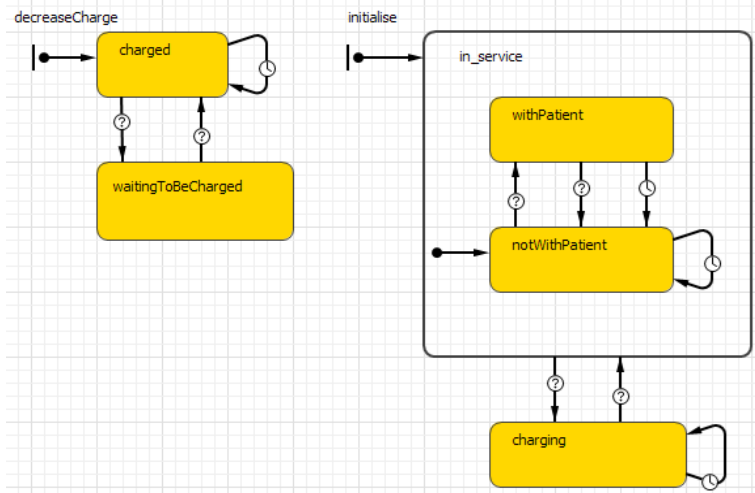

Figure 12: Robot agent

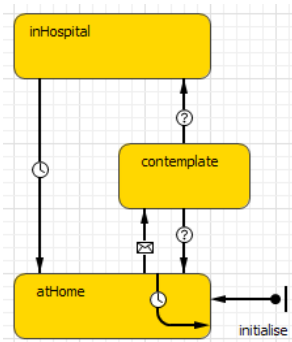

Figure 13: Visitor agent

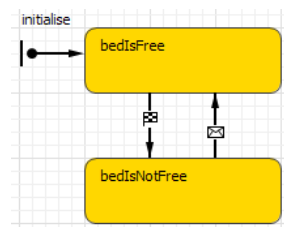

Figure 14: Bed agent

## Step 6: Defining Interactions

The figures below (Figure 15-17) show the most significant tasks and interactions of the simulation represented in sequence diagrams. I haven't transferred all interactions into sequence diagrams because there would be too many and it is not important to represent every interaction with a sequence diagram.

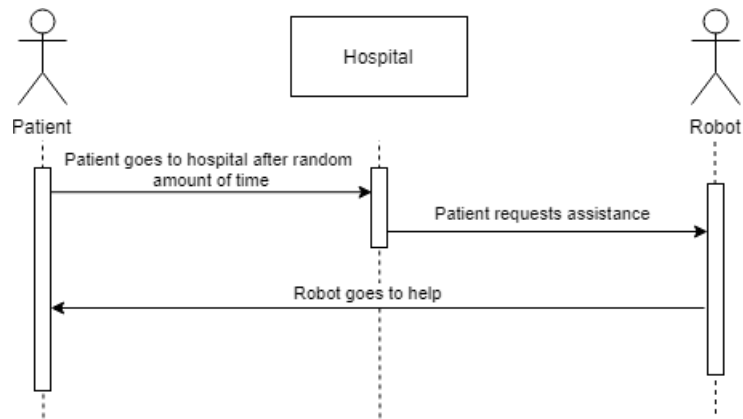

Figure 15: Patient getting treated by robot

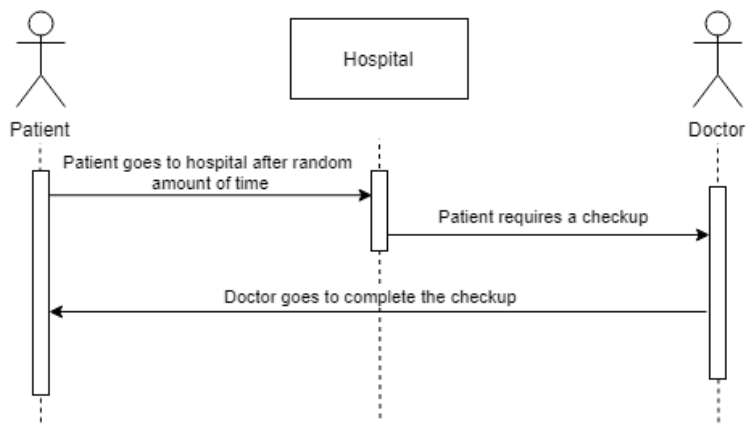

Figure 16: Patient getting treated by doctor

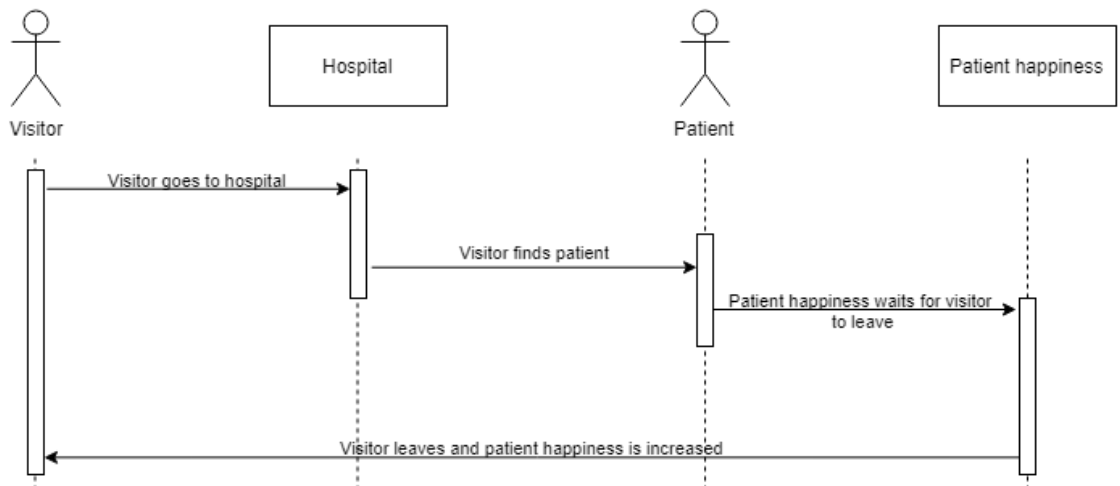

Figure 17: Visitor visits patient

## 5 The Final Version of the Hospital Simulation Model

Once the conceptual model was completed I have updated the simulation model to reflect all design features. Figure 18 shows a screenshot of the final hospital simulation model while running. The benefit of this final model in comparison to the previous versions created along the way is that this version is catered towards two of the hypotheses on the original conceptual model so, these being "Trust between machines and humans will affect human-human relationships in a negative way" and "Machine-look influences our decisions (trust)". This means that we can see valuable outputs that are related to the original ideas of the conceptual model e.g. the lines connecting each patient to other patients show how their trust values differ, if the line is red it means that those two patients' trust values differ a lot meaning their relationship is not good but if it is green it means their relationship is good. These colours are affected depending on lots of different aspects in this model that were not included in the previous, simpler models.

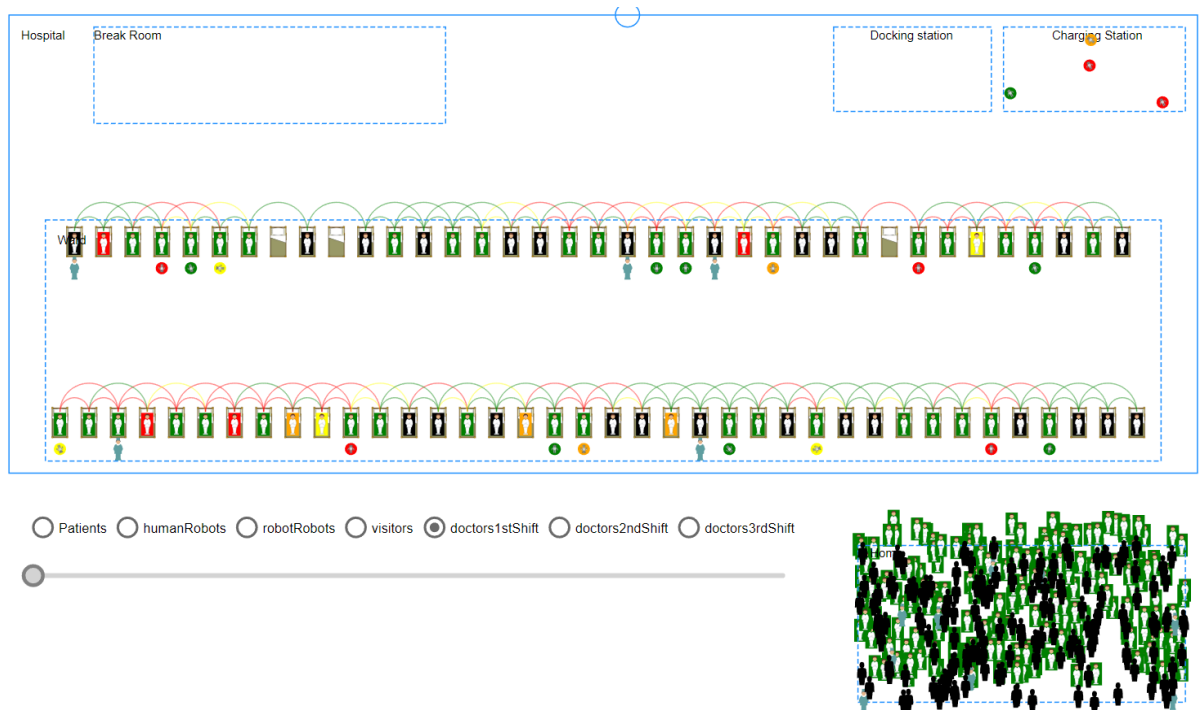

Figure 18: Screenshot of the final version of the simulation in progress

## 6 Conclusions

During my internship I finalised the conceptualisation of a simulation model, based on incomplete information previously gathered by Peer-Olaf Siebers and his colleagues, for exploratory studies of the impact of proactive machines in a hospital scenario. I have then implemented the simulation model, and documented my implementation using the EABSS framework.

The next step will be a more in-depth validation of the model. In particular we need to establish face validity by showing it to the people who provided us with the incomplete conceptual model. Once this has been done and the colleagues are happy with our interpretation of the information provided, we would need to run some experiments to test the hypotheses define in the conceptual model.

**NB:** The final version of the hospital simulation model is available upon request. Please get in touch with Peer-Olaf Siebers (peer-olaf.siebers@nottingham.ac.uk).

## Acknowledgement

The internship this working paper is based upon was run jointly by Peer-Olaf Siebers and Christian Wagner. The original (incomplete) conceptual model this work is based upon was donated by Peer-Olaf Siebers, Tommy Nilsson, and Elvira Perez Vallejos. Many thanks to all supporters!